\documentclass[12pt]{iopart}
\usepackage{iopams}
\usepackage{graphicx}
\input amssym.def \input amssym

\begin{document}

\title[]{Unitary reflection groups \\for quantum fault tolerance}

\author{Michel Planat$^{\dag}$ and Maurice Kibler$^{\ddag}$
}

\address{$^{\dag}$ Institut FEMTO-ST, CNRS, 32 Avenue de
l'Observatoire,\\ F-25044 Besan\c con, France }

\address{$^{\ddag}$Universit\'{e} de Lyon, F-69622, Lyon, France; Universit\'{e} Lyon 1, Villeurbanne; CNRS/IN2P3, UMR5822, IPNL}






\noindent
\hrulefill

\begin{abstract}

This paper explores the representation of quantum computing in terms of unitary reflections (unitary transformations that leave invariant a hyperplane of a vector space). The symmetries of qubit systems are found to be supported by Euclidean real reflections (i.e., Coxeter groups) or by specific imprimitive reflection groups, introduced (but not named) in a recent paper [Planat M and Jorrand Ph 2008, {\it J Phys A: Math Theor} {\bf 41}, 182001]. The automorphisms of multiple qubit systems are found to relate to some Clifford operations once the corresponding group of reflections is identified. For a short list, one may point out the Coxeter systems of type $B_3$ and $G_2$ (for single qubits), $D_5$ and $A_4$ (for two qubits), $E_7$ and $E_6$ (for three qubits), the complex reflection groups $G(2^l,2,5)$ and groups No $9$ and $31$ in the Shephard-Todd list. The relevant fault tolerant subsets of the Clifford groups (the Bell groups) are  generated by the Hadamard gate, the $\pi/4$ phase gate and an entangling (braid) gate [Kauffman L~H and Lomonaco S~J 2004 {\it New J. of Phys.} {\bf 6}, 134]. Links to the topological view of quantum computing, the lattice approach and the geometry of smooth cubic surfaces are discussed.

\end{abstract}


\section{Introduction}

Quantum computing is an exciting topic calling for a rich palette of mathematical concepts. Among them, group theory plays a considerable role being relevant for describing quantum errors (using Pauli and other error groups \cite{Kla02}) and quantum fault tolerance as well (using the Clifford group \cite{Clark07, Planat08}, the braid group \cite{Kitaev04,Kauf04} or the homological group \cite{Bombin07, Wirth08}). In this paper, we add to this list by showing the great relevance of real reflection groups (Coxeter groups), as well as unitary (complex) reflection groups, for representing a large class of protected quantum computations in a unifying geometrical language. 

Basically, a reflection in Euclidean space is a linear transformation of the space that leaves invariant a hyperplane while sending vectors orthogonal to the hyperplane to their negatives. Euclidean reflection groups of such {\it mirror symmetries} possess a {\it Coxeter group} structure, i.e., they are generated by a finite set of involutions and specific relations. More generally, unitary reflection groups (also known as groups of pseudo-reflections or complex reflection groups) leave a hyperplane pointwise invariant within the complex vector space \cite{Hump90, Kane01}. The simplest example of a Coxeter group is the dihedral group $\mbox{Dih}_n$ ($n>2$), which is the symmetry group of a regular polygon with $n$ vertices/edges: it is easy to visualize that $\mbox{Dih}_n$ consists of $n$ rotations (through multiples of $2\pi/n$) and $n$ reflections (about the {\it diagonals} of the polygon)\footnote{In the Schoenflies notation of molecular physics, the group $\mbox{Dih}_n$ is denoted $C_{nv}$ or $D_n$ according to wheather as it is realized in terms of proper and improper rotations or proper rotations only, repectively.}. The symmetry group of a regular $n$-simplex (a $1$-simplex is a line segment, a $2$-simplex is a triangle and a $3$-simplex is a tetrahedron) is the symmetric group $S_{n+1}$, also known as the Coxeter group of type $A_n$.

To motivate our approach, let us mention that those two types of groups immediately appear for qubits. The dihedral group $\mbox{Dih}_4$ (corresponding to  the set of symmetries of the square) is the group of automorphisms of a pair of  observables taken in the Pauli group $\mathcal{P}_1=\left\langle \sigma_x, \sigma_y, \sigma_z\right\rangle$, generated by the Pauli matrices and, among other instances, $\mbox{Dih}_{6}$ (corresponding to the set of symmetries of the hexagon) is the group of outer automorphisms of $\mathcal{P}_1$. The symmetry group $S_4$ of the tetrahedron is known to be relevant in the optimal qubit tomography based on the Bloch sphere \cite{Durt08}. As we will see below, $S_4$ is also hidden in the (less trivial) Coxeter group $B_3=\mathbb{Z}_2 \times{S_4}$ (associated with the {\it snub cube}), the group of symmetries of all the automorphisms of $\mathcal{P}_1$. It is well known that all symmetry groups of regular polytopes are finite Coxeter groups. Finite Coxeter groups either belong to four infinite series $A_n$, $B_n$, $D_n$ and $I_2(n)$ ($n$-gon), or are of the exceptional type $H_3$ (the icosahedron/dodecahedron), $F_4$ (the $24$-cell), $H_4$ (the $120$-cell/$600$-cell), $E_6$, $E_7$ and $E_8$ (associated with the  polytopes of the same name). The type $E_7$ was recently proposed as a candidate to model quantum entanglement in analogy to the entropy of BPS black holes \cite{Levay07}. 

A Coxeter group arises in a simple Lie algebra as the Weyl group attached to the root system of the algebra.  Specifically, the Weyl-Coxeter group for a given simple Lie algebra is generated by reflections through the hyperplanes orthogonal to the roots \cite{Carter89}. Not all Coxeter groups appear as Weyl groups of a Lie algebra, because some of them lack the property of being {\it crystallographic}, a distinctive feature of some root systems. Not just orthogonal reflections leaving invariant a hyperplane passing through the origin can be defined. One can also define the affine Weyl group, composed of affine reflections relative to a {\it lattice} of hyperplanes. The lattices left invariant by some affine (and crystallographic) Weyl group also exist in four infinite series denoted $A_n\widetilde{}$, $B_n\widetilde{}$, $C_n\widetilde{}$ and $D_n\widetilde{}$, and there are six exceptional types. Affine Weyl groups are infinite Coxeter groups that contain a normal abelian subgroup such that the corresponding quotient group is finite and is a Weyl group. To pass from the finite Coxeter graph to the infinite one, it suffices to add an additional involution and one or two additional relations. May be the best illustrative example is the way from the hexagon to the hexagonal tiling. The Coxeter group $\mbox{Dih}_6$ (corresponding to the Coxeter system $I_2(6)$, also called $G_2$) is represented by two generators $x_1$ and $x_2$ such that that $x_1^2=x_2^2=(x_1 x_2)^6=1$. The corresponding Coxeter graph contains two vertices and one edge indexed with the integer $6$.
The Coxeter group of the hexagonal tiling is obtained by adding one involution $x_3$ and two extra relations, viz, $x_3^2=(x_2x_3)^3=(x_1x_3)^2=1$. The hexagonal lattice reminds us of the geometry of graphene quantum dots, which were recently proposed for creating coherent spin qubits \cite{Trauz07}.

Let us pass to the unitary reflection groups. The irreducible ones were classified \cite{Shep54} and found to form an infinite family $G(m,p,n)$ (with $p$ dividing $m$) and $34$ exceptional cases. The infinite family contains the infinite families of finite Coxeter groups as special cases. In particular, $G(n,n,2):=I_2(n)$. In our recent paper \cite{Planat08}, we arrived at the conclusion that the automorphisms of sets of mutually unbiased bases for multiple qubits are controlled by the groups $\mathbb{Z}_2^l \wr A_5$, in which $A_5$ is the alternating group on five symbols and $\wr$ is the wreath product, i.e., the semi-direct action of the permutation group $A_5$ on five copies of the two-element group $\mathbb{Z}_2$. It was not recognized at that time that those groups are precisely the Coxeter groups of systems $G(2^l,2,5)$, with the special case $W(D_5)=G(2,2,5)$ (defining the Weyl group $\mathbb{Z}_2 \wr A_5$) corresponding to the two-qubit system. The relation between Clifford groups, unitary reflection groups and coding theory was studied in Ref \cite{Nebe01}.

All these remarkable relationships between the symmetries of qubit systems and reflection groups are the origin of our motivation to undertake a parallel between the properties of Coxeter systems and quantum coherence. A further support to this idea is that, to any unitary reflection group, one can associate a generalized braid group \cite{Broue00}. Braid groups, which play an important role in anyonic symmetries, already paved their way in the quantum computing literature \cite{Kauf04}. In the present paper, our goal is to establish some new bridges between the geometry of groups, encoded into the reflection groups, and quantum information processing tools. 

This paper is organized as follows. In Sec 2 we provide a technical introduction to reflection groups, with specific examples relevant for the present paper. In Sec 3 we remind of some recently established links between finite geometries and the observables of multiple qubit systems \cite{Pauligraphs}. The corresponding automorphism groups are derived and a representation in terms of finitely presented groups of reflections is displayed whenever possible. The irruption of imprimitive groups of type $G(2^l,2,5)$ for representing the symmetries of complete sets of mutually unbiased bases is explained. In Sec 4 we recall some useful concepts of group extensions used to address topics such as Clifford groups and their relations to error groups. A particular entangling subgroup of the two- and three-qubit Clifford groups is exhibited and its relation to topological quantum computation, the Yang-Baxter equation and the Coxeter system of type $D_5$ and $E_6$ is discussed. Finally, smooth cubic surfaces are evoked to vindicate this connection.     

\section{A primer on reflection groups and root systems}
\subsection{Reflections}

To begin with, let us start with an $l$-dimensional (real) Euclidean space $\mathbb{E}$, endowed with a product $(.,.)$ such that $\forall a,b \in \mathbb{R}$ and $\forall x,y \in \mathbb{E}$, we have  $(x,y)=(y,x)$ (symmetry),  $(ax+by,z)=a(x,z)+b(y,z)$ (linearity), $(x,x)\ge 0$ and $(x,x)=0$ $\Rightarrow$ $x=0$ (a positive definite form). Let us introduce the orthogonal group $O(\mathbb{E})$ of linear transformations $f$ of $\mathbb{E}$ as 
%
$$O(\mathbb{E})=\left\{f:\mathbb{E}\rightarrow \mathbb{E}|\forall x,y \in \mathbb{E}:(f(x),f(y))=(x,y)\right\}.\nonumber$$

%
Let $H_{\alpha}\subset \mathbb{E}$ be the hyperplane
%
$$H_{\alpha}=\left\{x|(x,\alpha)=0\right\},\nonumber$$
%
then a reflection $s_{\alpha}:\mathbb{E}\rightarrow \mathbb{E}$ is defined as
%
$$s_{\alpha}(x)=x~\mbox{if}~x\in H_{\alpha}~\mbox{and}~ s_{\alpha}(\alpha)=-\alpha.\nonumber$$
%
It is clear that $s_{\alpha}\in O(\mathbb{E})$, i.e., $(s_{\alpha}(x),s_{\alpha}(y))=(x,y)$. There are two further important properties

(i) The reflection $s_{\alpha}(x)$ of each vector $x\in \mathbb{E}$ can be explicitely defined using the action of the linear product
%
$$\forall x \in \mathbb{E}:s_{\alpha}(x)=x-2\frac{(x,\alpha)}{(\alpha,\alpha)}\alpha.\nonumber$$
%

(ii) Let $t\in O(\mathbb{E})$. An hyperplane maps to a hyperplane under the action of $t$:
%
$$t(H_{\alpha})=H_{t(\alpha)},\nonumber$$
%
and a reflection maps to a reflection under conjugation in $O(\mathbb{E})$:
%
$$t s_{\alpha}t^{-1}=s_{t(\alpha)}.\nonumber$$
%
Given $W\subset O(\mathbb{E})$, $W$ is a Euclidean reflection group if $W$ is generated, as a group, by reflections. It is irreducible if it cannot be rewritten as a product of two reflection groups.

\subsection{Root systems}
\label{roots}

The concept of a finite Euclidean reflection group may be reformulated in terms of linear algebra by using its {\it root system} $\Delta$.

For doing this, one replaces each reflecting hyperplane of the reflection group $W$ by its two orthogonal vectors of unit length. Let $\Delta \subset \mathbb{E}$ be the resulting set of vectors. The vectors of $\Delta$ satisfy two important properties

(I) If $\alpha \in \Delta$, then $\lambda\alpha \in \Delta$ iff $\lambda=\pm 1$.

(II) The set $\Delta$ is permuted under the action of $W$: If $\alpha,\beta \in \Delta$, then $s_{\alpha}(\beta)\in\Delta$.

Any element of $\Delta$ is a {\it root}, and $\Delta$ is named a {\it root system}.

It is noteworthy that among root systems are those that possess the extra property of being {\it crystallographic}. Besides (I) and (II), such systems satisfy

%
$\mbox{(III)}$ For any $\alpha,\beta \in \Delta$, one has $\left\langle \alpha,\beta\right\rangle:=2\frac{(\alpha,\beta)}{(\alpha,\alpha)}\in \mathbb{Z}.$
%

The reflection groups having a crystallographic root system are called Weyl groups and the integers $\left\langle \alpha,\beta\right\rangle$ in (III) are called Cartan integers. The matrix whose elements are the Cartan integers is called the Cartan matrix. Crystallographic groups arise in the context of semi-simple complex Lie algebras as an intrinsic property of the symmetries of their roots \cite{Carter89}, and it is precisely in this context that their classification was established \footnote{In III, the notation $\left\langle \alpha,\beta\right\rangle$ for denoting the integers occuring in the crystallographic root system should not be confused with the bra/ket Dirac notation used in quantum mechanics when dealing with a Hilbert space formalism. The brackets are useful for comparing the roots $\alpha$ and coroots $\alpha^\vee=2\frac{\alpha}{(\alpha,\alpha)}$ thanks to the relation $\left\langle \alpha,x\right\rangle =(\alpha^\vee,x)$. The bracket notation is also conventionaly used for the finite presentation of a group (as in Sec \ref{Coxeter} and elsewhere).}.

\subsection{Coxeter systems}
\label{Coxeter}

The algebraic structure of finite Euclidean reflection groups can be understood via the concept of a Coxeter system. It can be used to classify finite reflection groups.

A group $W$ is a {\it Coxeter group} if it is finitely generated  by a subset $S\subset W$ of involutions and pairwise relations
\begin{equation}
W=\left\langle s\in S|(ss')^{m_{ss'}}=1\right\rangle,
\end{equation}
where $m_{ss}=1$ and $m_{ss'}\in\left\{2,3,\ldots\right\}\cup\left\{\infty \right\}$ if $s\neq s'$. The pair $(W,S)$ is a Coxeter system, of rank $|S|$ equal to the number of generators.  One can associate a Coxeter system to any finite reflection group.

Coxeter systems are conveniently represented by Coxeter graphs. A {\it Coxeter graph} $X$ is a graph with each edge labelled by an integer $\ge 3$. The standard method of assigning a Coxeter graph to a Coxeter system $(W,S)$ is as follows: (i) $S$ gives the vertices of $X$, (ii) given $s,s'\in S$ there is no edge between $s$ and $s'$ if $m_{ss'}=2$, (iii) given $s,s'\in S$ there is an edge labelled by $m_{ss'}$ if $m_{ss'}\ge 3$. This assignment sets up a one-to-one correspondence between a Coxeter system and its associated Coxeter graph.  

Let us illustrate the above concepts with examples pertaining to quantum computing. The Coxeter system of type $G_2=I_2(6)$ controls the outer automorphisms of the Pauli group (see Sec \ref{singlequbit}). As already announced in the introduction, its presentation immediately follows from the one of a rank $n$ dihedral group

$$\mbox{Dih}_n=\left\langle s_1,s_2|(s_1)^2=(s_2)^2=(s_1s_2)^n=1\right\rangle.$$

The Coxeter system $A_3$ of Weyl group $S_4$ appears in the tomography of qubits \cite{Durt08}. It is of rank three with representation
$$S_3=\left\langle s_1,s_2,s_3|(s_1)^2=(s_2)^2=(s_3)^2=(s_1s_3)^2=(s_1s_2)^3=(s_2s_3)^3=1\right\rangle.$$

Coxeter systems of the type $D_5$ and of exceptional type $E_6$ and $E_7$ occur in topological quantum computing. For a finite representation of $E_6$, see Eq \ref{presE6}. 
 
\subsection{Fundamental root systems as Coxeter systems}

The equivalence between finite reflection groups and Coxeter systems follows from the introduction of fundamental root systems. Given a root system $\Delta \subset \mathbb{E}$, then $\Sigma \subset \Delta$ is a {\it fundamental system} of $\Delta$ if (i) $\Sigma$ is linearly independant, (ii) every element of $\Delta$ is a linear combination of elements of $\Sigma$ where the coefficients are all non-negative or all non-positive. The elements of $\Sigma$ are called the {\it fundamental roots}. It can be shown that there is a unique fundamental system $\Sigma$ associated with any root system $\Delta$ of a finite reflection group.

To a fundamental root $\alpha \in \Sigma$, there is associated a fundamental reflection $s_{\alpha}$. Furthermore, given the fundamental system $\Sigma$ of $\Delta$, then $W(\Delta)=W$ is generated by fundamental reflections $s_{\alpha}$. We want to associate a bilinear form (see Sec 2.1) to every Coxeter system. Define the bilinear form $\mathcal{B}:\Sigma \times \Sigma \rightarrow \mathbb{R}$ by 

$$\mathcal{B}(\alpha_s,\alpha_{s'})=-\mbox{cos}(\frac{\pi}{m_{s s'}}).$$

In particular, $\mathcal{B}(\alpha_s,\alpha_s)=1$ and $\mathcal{B}(\alpha_s,\alpha_{s'})=0$ when $m_{ss'}=2$. The bilinear form can be shown to be positive definite for every finite Coxeter system $(W,S)$. It may be identified with the original inner product in $\mathbb{E}$. 

Given a fundamental system $\Sigma=(\alpha_1,\ldots,\alpha_l)$ of $\Delta$, we then assign a Coxeter graph $X$ to $\Delta$ by the rules 

(i) $\Sigma$ gives the vertices of $X$.

(ii) Given $\alpha_i \neq \alpha_j \in \Sigma$, there is no edge between $\alpha_i$ and $\alpha_j$ if $m_{ij}=2$ (i.e., $\alpha_i$ and $\alpha_j$ are at right angles).

(iii) Given $\alpha_i\neq \alpha_j \in \Sigma$, there is an edge labelled by $m_{ij}$ if $m_{ij}\ge 3$.

As a result, the Coxeter graph of root system $\Delta$ is the Coxeter graph of the Weyl group $W(\Delta)$.

Let see how it works for the examples listed in Sec 2.3 above. To the dihedral group $\mbox{Dih}_n$ is associated the root system
$$G_2(m)=\left\{\left(\cos(\frac{k\pi}{m}),\sin(\frac{k\pi}{m})\right)|0\le k\le 2m-1\right\},$$
with $\alpha_1=(\mbox{cos}(\frac{\pi}{m}),\mbox{sin}(\frac{\pi}{m}))$ and $\alpha_2=(\mbox{cos}(\frac{2\pi}{m}),(\mbox{sin}(\frac{2\pi}{m}))$.

To the symmetric group $S_{l+1}$ is associated the root system
$$A_1=\left\{\epsilon_i-\epsilon_j|i\neq j,~1\leq i,~j\leq l+1\right\}$$
and the fundamental system
$$\Sigma=\left\{\epsilon_1-\epsilon_2, \epsilon_2-\epsilon_3,\ldots,\epsilon_l-\epsilon_{l+1}\right\} ,$$
in which $\left\{\epsilon_j\right\}$ is an orthonormal basis of $\mathbb{R}^{l+1}$.

For an exhaustive list of root systems, see \cite{Kane01}, p 93.

\subsection{The weight lattice of a Weyl group}

As already stressed in Sec \ref{roots}, a Weyl group is a reflection group  satisfying the crystallographic axiom III. Furthermore, to every Weyl group $W$ one can associate a lattice of integers which is stabilized by the action of $W$ on the roots. Let us define the weight lattice $\mathcal{L}_W$ by
$$\mathcal{L}_W=\left\{x\in \mathbb{E}|\forall \alpha \in \Delta:\left\langle \alpha,x\right\rangle \in \mathbb{Z} \right\}.$$
Conversely, the Weyl group is uniquely determined by its weight lattice $\mathcal{L}_W$.

For instance, we obtain
$$\mathcal{L}_{W(I_2(4))}=\left(\begin{array}{cc} 2 & 1 \\3	 & 2\end{array}\right),~~~~\mathcal{L}_{W(A_2)}=\left(\begin{array}{cc} 2 & 1 \\1	 & 2\end{array}\right).  $$
A further example  is given at the end of Sec \ref{tolerance}.

\subsection{Affine Weyl groups}

One can start from the Weyl group of a crystallographic root system and form an infinite group that still possesses a structure analogous to that of the Weyl group, i.e., a Coxeter group structure. Hyperplanes are defined as
$$H_{\alpha,k}=\left\{t\in \mathbb{E}|(\alpha,t)=k\right\},$$
and the reflection $s_{\alpha,k}$ through the hyperplane $H_{\alpha,k}$ reads
$$s_{\alpha,k}(x)=x-(\alpha,x)\alpha^{\vee}+k\alpha^{\vee},~\mbox{with}~\mbox{coroot}~\alpha^{\vee}=2\frac{\alpha}{(\alpha,\alpha)}.$$
By definition, the {\it affine Weyl group} $W_{\mbox{aff}}(\Delta)$ is generated by the set of reflections $\left\{s_{\alpha,k}|\alpha \in \Delta,~k\in \mathbb{Z}\right\}$. For details and the classification of affine Weyl groups, see \cite{Kane01}, p 101.
 
\subsection{Unitary reflection groups}

Euclidean reflection groups may be generalized as pseudo-reflection groups by replacing the real Euclidean space by an arbitrary vector space over a field $\mathbb{F}$. We shall mention complex reflection spaces, defined over the complex field $\mathbb{C}$, which we shall use later for protected qubits. 
 
Rather than an inner product, we shall use a positive definite Hermitian form $(.,.)$ acting on a complex finite-dimensional vector space $V$. Every reflection $s:V\rightarrow V$ of order $n$ over $\mathbb{C}$ satisfies the reflection property
$$s(x)=x+(\xi -1)\frac{(\alpha,x)}{(\alpha,\alpha)}\alpha,$$
for all $x \in V$, where $\xi$ is a primitive $n$-th root of unity, $\alpha$ is an eigenvector such that $s(\alpha)=\xi \alpha$ and $(x,y)$ is a positive definite Hermitian form satisfying $(s(x),s(y))=(x,y)$.

Finite irreducible unitary reflection groups were classified \cite{Shep54}. They consist of three infinite families $\left\{\mathbb{Z}/m\mathbb{Z}\right\}$, $\left\{S_n\right\}$, $\left\{G(m,p,n)\right\}$, and $34$ exceptional cases that we denote $\mathcal{U}_n$, $n=1..34$ (see \cite{Kane01}, p 161). 
We shall be concerned with {\it imprimitive unitary reflection groups}. A group $G\subset GL(V)$ is said to be {\it imprimitive} if there exists a decomposition $V=V_1\otimes \ldots \otimes V_k$ ($k\ge 2$), where the subspaces $V_i$ are permuted transitively by $G$. If $p|m$, we can define the semidirect group
$$G(m,p,n)=A(m,p,n)\rtimes S_n ,$$
where the permutation group $S_n$ is isomorphic to a subgroup of $GL_n(\mathbb{C})$ and
$$A(m,p,n)=\left\{\mbox{Diag}(\omega_1,\omega_2,\ldots,\omega_{n-1},\omega_n)|\omega_i^m=1~\mbox{and}~(\omega_1\ldots\omega_n)^{m/p}=1 \right\}.$$

Many Euclidean  reflection groups are special cases, including
$$G(1,1,n)=S_n=W(A_{n-1}),$$

$$G(m,m,2)=\mathbb{Z}/m\mathbb{Z}\rtimes S_2=\mbox{Dih}_m=W(I_2(m)),$$

$$G(2,2,n)=(\mathbb{Z}/m\mathbb{Z})^{n-1}\rtimes S_n=W(D_n).$$

We shall be concerned later with a generalisation $G(2^l,2,5)$ of the $D_5$ Coxeter system (see Sec \ref{MUBs}).

\section{Automorphisms of multiple qubit systems as reflection groups}

\subsection{The single qubit case}
\label{singlequbit}

In the sequel of the paper, we use several important concepts of group theory such as normal subgroups, short exact sequences and automorphism groups. A reminder can be found in Appendix 1. In this section, the link between quantum error groups and reflection groups is studied. Most often, in the quantum computing context, tensor products of Pauli matrices (for $\frac{1}{2}$-spin) are considered as error groups \cite{Clark07,Planat08}. We shall denote $\mathcal{P}_n$ the $n$-qubit Pauli group \footnote{The $n$-qubit Pauli group is in general not isomorphic to the single qubit Pauli group in dimension $2^n$. The latter group is most often denoted as the Heisenberg-Weyl group \cite{Kibler08}}, obtained by taking tensor products of $n$ ordinary Pauli matrices up to a phase factor $Z(\mathcal{P}_n)=\left\{\pm 1, \pm i\right\}$. Symmetries underlying $\mathcal{P}_n$, i.e., automorphisms of $\mathcal{P}_n$ are related to reflection groups. 
Other relations to reflection groups arise in quantum error-correcting codes and Clifford groups, as shown in the next section. Many of our calculations make use of the group theoretical packages GAP \cite{GAP} and Magma \cite{MAGMA}. 

Let us start with the single qubit case for which our claim takes a very simple form, already advertized in the introduction. The single qubit Pauli group $\mathcal{P}_1$ is generated by the Pauli spin matrices $\sigma_0$ (the identity matrix), $\sigma_x$ (the shift matrix), $\sigma_z$ (the flip matrix) and $\sigma_y=i\sigma_x\sigma_z$. It is of order $16$ and may be identified to the imprimitive reflection group $G(4,2,2)$. The group of automorphisms of $\mathcal{P}_1$ is 
\begin{equation}
\mbox{Aut}(\mathcal{P}_1)\cong \mathbb{Z}_2^3 \rtimes S_3=W(B_3)\cong \mathbb{Z}_2 \times S_4=W(A_1 A_3),
\end{equation}  
in which $\mathbb{Z}_2=\mathbb{Z}/2\mathbb{Z}$ and \lq\lq$\times$" and \lq\lq $\rtimes$" are, respectively, a direct and semi-direct product. 
The Coxeter group $W(B_3)$ corresponds to the first description $\mathbb{Z}_2^3 \rtimes S_3$, but one can also use the second description $\mathbb{Z}_2 \times S_4$ to produce the Weyl group of a reducible Coxeter system of type $A_1 A_3$ and rank 4. 
The generating relations of the irreducible Coxeter system $B_3$ are
\begin{equation}
x_1^2=x_2^2=x_3^2=(x_1x_2)^3=(x_2x_3)^4=(x_1x_3)^2=1.
\end{equation}  
In the Wenninger classification of polyhedron models \cite{Wenn79}, the symbols $W_1$ to $W_5$ correspond to platonic solids (the regular polyhedra), $W_6$ to $W_{18}$ to Archimedean solids (the semi-regular polyhedra), the remaining ones go from $W_{19}$ to $W_{119}$. The {\it snub cube} corresponds to the symbol $W_{17}$ and its automorphism group is the Coxeter group $W(B_3)$. It comprises $38$ faces, of which $6$ are squares and other $32$ are equilateral triangles. 

Inner automorphisms of $\mbox{Inn}(\mathcal{P}_1)$ form a normal subgroup of $\mbox{Aut}(\mathcal{P}_1)$ isomorphic to $\mathbb{Z}_2 \times \mathbb{Z}_2$. The outer automorphism group $\mbox{Aut}(\mathcal{P}_1)/\mbox{Inn}(\mathcal{P}_1)$ reads
\begin{equation}
\mbox{Out}(\mathcal{P}_1)\cong \mbox{Dih}_6=W(G_2)\cong \mathbb{Z}_2 \times \mbox{Dih}_3=W(A_1 I_2(3)).
\label{Out2}
\end{equation}  
It can be represented by the irreducible (rank $2$) Coxeter group $G_2$ or by a reducible Coxeter system of rank $3$ composed of its two factors $A_1$ and $I_2(3)$. It is most surprising that $\mbox{Out}(\mathcal{P}_1)$ and $\mbox{Aut}(\mathcal{P}_1)$ are the academic examples treated in \cite{Kane01} (p 67).

The generating relations of the Coxeter group $W(G_2)$ are
\begin{equation}
x_1^2=x_2^2=(x_1x_2)^6=1.
\end{equation}  
They correspond to the symmetries of the hexagon. It may be useful to mention that there does not exist a one-to-one relation between a group and its automorphism group, or between a group and its outer automorphism group. In the present case one observes that the simple group $M_{21}=PSL(3,4)$ has $\mbox{Dih}_6$ as its outer automorphism group ($M_{21}$ is not a Coxeter group but a group of Lie type \cite{Carter89}). The group $M_{21}$, of order $20160$, is the stabilizer of a point in the large Mathieu group $M_{22}$, defined from the Steiner system \footnote{A Steiner system $S(a,b,c)$ with parameters $a$, $b$, $c$, is a $c$-element set together with a set of $b$-element subsets of $S$ (called {\it blocks}) with the property that each $a$-element subset of $S$ is contained in exactly one block. A finite projective plane of order $q$, with the lines as blocks, is an $S(2, q+1, q^2+q+1)$, because it has $q^2+q+1$ points, each line passes through $q+1$ points, and each pair of distinct points lies on exactly one line.} $S(3,6,22)$, and the stabilizer of a triad in the Mathieu group $M_{24}$.  This comment is written in relation to the occurence of Mathieu group $M_{22}$, as well as $M_{20}=W(D_5)$ within the context of two-qubit systems (see \cite{Planat08} and Secs \ref{MUBs} and \ref{topological}).

Let us pass to the other types of reflection groups, which may be associated with single qubits. One may wish to define a reflection group from the outer automorphisms at each location of a {\it lattice} reflection group. As announced in the introduction, to the finite Coxeter group $G_2$ corresponds the affine Weyl group $H_2\widetilde{}$ (also called $G_3$), a rank three (infinite) reflection group, with the following generating relations 
\begin{equation}
x_1^2=x_2^2=x_3^2=(x_1x_2)^3=(x_2x_3)^6=(x_1x_3)^2=1.
\end{equation}  
It is associated with a hexagonal (or triangular) tesselation of the plane. The most relevant qubit model may well be a cluster state model \cite{Nest08}, and one may want to think about graphene as a possible real world realization. 

\subsection{The two-qubit case}

The two-qubit Pauli group $\mathcal{P}_2$ is more involved than $\mathcal{P}_1$. In particular, it features entangled states. There exists in-depth studies of them in the quantum information literature, but the present approach is performed in the spirit of \cite{Pauligraphs}. The two-qubit Pauli group may be generated as $\mathcal{P}_2=\left\langle\sigma_0 \otimes \sigma_x,\sigma_x \otimes \sigma_x,\sigma_z \otimes \sigma_z,\sigma_y \otimes \sigma_z,\sigma_z \otimes \sigma_x \right\rangle$. It is of order $64$. The group of automorphisms of $\mathcal{P}_2$ was already featured in \cite{Planat08}   
\begin{equation}
\mbox{Aut}(\mathcal{P}_2)\cong U_6.\mathbb{Z}_2^2 ~~\mbox{with}~U_6\cong \mbox{Aut}(\mathcal{P}_2)'=\mathbb{Z}_2^4 \rtimes A_6
\label{U6}
\end{equation}  
and \lq\lq." means that the short exact sequence $1\rightarrow U_6 \rightarrow \mbox{Aut}(\mathcal{P}_2)\rightarrow \mathbb{Z}_2^2 \rightarrow 1$ does not split. Neither $\mbox{Aut}(\mathcal{P}_2)$ nor $U_6$ are Coxeter groups. The dividing line between $\mbox{Aut}(\mathcal{P}_2)$ and a Coxeter group may be  appreciated by displaying the Weyl group for Coxeter system $B_6$, of the same cardinality, which may be written as the semidirect product $W(B_6)=\mathbb{Z}_2 \wr S_6=\mathbb{Z}_2^6 \rtimes S_6$. 

The group $U_6$ is an important maximal subgroup of several sporadic groups. The group of smallest size where it appears is the Mathieu group $M_{22}$. 
Mathieu groups are sporadic {\it simple} groups, so that $U_6$ cannot be normal in $M_{22}$. It appears in a subgeometry of $M_{22}$ known as a {\it hexad}. 

Any large Mathieu group can be defined as the automorphism (symmetry) group of a Steiner system \cite{Wilson}. The group $M_{22}$ stabilizes the Steiner system $S(3,6,22)$ comprising $22$ points with $6$ points in any block, each set of $3$ points being contained exactly in one block. Any block in $S(3,6,22)$ is a Mathieu hexad, i.e., it is stabilized by the group $U_6$. There exists up to equivalence a unique S(5,8,24) Steiner system called a Witt geometry. The group $M_{24}$ is the automorphism group of this Steiner system, that is, the set of permutations which map every block to some other block. The subgroups $M_{23}$ and $M_{22}$ are defined to be the stabilizers of a single point and two points respectively.

The outer automorphism group of the two-qubit Pauli group 
\begin{equation}
\mbox{Out}(\mathcal{P}_2)\cong \mathbb{Z}_2 \times S_6=W(A_1A_5)
\label{Out1}
\end{equation}  
corresponds to the reducible Coxeter system $A_1 A_5$\footnote{ The Coxeter system $A_5$ should not be confused with the alternating group $A_5$. The meaning of $A_5$ should be clear from the context.}.

\subsection{Automorphisms of central quotients of Pauli groups}

We failed to discover a general rule for the automorphism group of the multiple qubit Pauli group $\mathcal{P}_n$. But there exists a very simple formula for the automorphism group of the central quotient $\tilde{\mathcal{P}}_n \cong \mathbb{Z}_2^{2n}$. It is easy to check that $\mbox{Aut}(\tilde{\mathcal{P}}_1)=\mathbb{Z}_6$, $\mbox{Aut}(\tilde{\mathcal{P}}_2)=A_8 \cong PSL(4,2)$ (of order $20160$), $\mbox{Aut}(\tilde{\mathcal{P}}_3)=PSL(6,2)$ (of order $20~158~709~760$). All automorphisms are found to be outer. More generally 
\begin{equation}
\mbox{Aut}(\tilde{\mathcal{P}}_n)\cong PSL(2n,2)=A_{2n-1}(2).
\end{equation}  
The group $PSL(2n,2)$ is the group of Lie type $A_{2n-1}$ over the field $\mathbb{F}_2$ \cite{Carter89}. For $PSL(2n,2)$, the Weyl group is the one defined by the Coxeter system of type $A_{2n-1}$, i.e., the symmetry group $S_{2n}$. 
The group $PSL(2n,2)$ also corresponds to the automorphism group of the ($n-1$)-qubit CSS (Calderbank-Schor-Steane) {\it additive} quantum code \cite{Nielsen2000}. The five-qubit Schor code and the seven-qubit Steane code have automorphism group $A_8$ and $PSL(6,2)$, respectively.  
 
\subsection{Geometric hyperplanes of the two-qubit system and their automorphism group}
\label{hyper}

This section is of slightly different flavour than the rest of the paper. It makes use of the finite geometries embodied by the commutation relations of observables within the Pauli group $\mathcal{P}_2$. Commuting/anti-commuting relations between the Pauli operators of the two-qubit system have been determined \cite{Pauligraphs}. They have been found to form the {\it generalized quadrangle} of order two $\mbox{GQ}_2$ and to admit three basic decompositions in terms of {\it geometric hyperplanes}. It is our purpose here to explicit how the outer automorphisms of such structures relate to Coxeter groups. 

A finite geometry is a set of points and lines together with incidence axioms. A {\it generalized quadrangle} $GQ$ obeys the following axioms: (i) It is a near-linear space, i.e., a space of points and lines such that any line has at least two points and two points are on at most one line, (ii) given an antiflag (a line  and a point not on the line) there is exactly one line through the point that intersects the line at some other point. A $\mbox{GQ}$ is said to be of order $(s,t)$ if every line contains $s+1$ points and every point is in exactly $t+1$ lines. The $\mbox{GQ}$ is called {\it thick} if both $s$ and $t$ are larger than $1$. If $s=t$, we simply speak of a $\mbox{GQ}$ of order $s$, that we denote $\mbox{GQ}_s$. The smallest thick generalized quadrangle $\mbox{GQ}_2$ contains $15$ points and $15$ lines, the axioms are dual for points and lines . A {\it geometric hyperplane} of a finite geometry is a set of points such that every line of the geometry either contains exactly one point of the hyperplane, or is completely contained in it. For $\mbox{GQ}_2$, there are three types of hyperplanes: a {\it perp-set}, a {\it grid} and an {\it ovoid} \cite{Pauligraphs}. The group of automorphisms of $\mbox{GQ}_2$ is the symmetric group $S_6$. (For the occurence of the generalized quadrangle $\mbox{GQ}_3$ see the end of Sec \ref{topological}.)  

%
%

Let us see now how finite geometries connect with the two-qubit system. Let us consider the fifteen tensor products $\sigma_i\otimes\sigma_j$ of ordinary Pauli matrices $\sigma_i \in \left\{\sigma_0,\sigma_x,\sigma_y,\sigma_z\right\}$, label them as follows $1=\sigma_0\otimes \sigma_x$, $2=\sigma_0\otimes \sigma_y$, $3=\sigma_0\otimes \sigma_z$, $a=\sigma_x \otimes \sigma_0$, $4=\sigma_x \otimes \sigma_x,\ldots$, $b=\sigma_y \otimes \sigma_0,\ldots$, $c=\sigma_z \otimes I_2,\ldots$, $15=\sigma_z \otimes \sigma_z$. One may take a point as an observable of the above set, and a line as a maximal set of mutually commuting operators, so that the geometry of $\mbox{GQ}_2$ is reproduced. 

In Eq (\ref{Out1}), we established that the observables in the Pauli group $\mathcal{P}_2$, which also span $\mbox{GQ}_2$, possess outer automorphisms forming the Weyl group of the reducible Coxeter system $A_1A_5$. We now intend to check if the observables spanning the hyperplanes of $\mbox{GQ}_2$ still have automorphisms controlled by some Coxeter system. Let us list the three hyperplanes $H_1$, $H_2$ and $H_3$ considered in Sec (3) of \cite{Pauligraphs} 
.

1) A {\it perp-set} $H_1$ of $\mbox{GQ}_2$, of cardinality $7$, is is defined by three lines passing through the reference point $a$, one can choose $H_1=\left\{(1,a,4),(2,a,5),(3,a,6)\right\}$. None of the lines  of $H_1$ carries an entangled state, observables in each of the lines form the group $\mathbb{Z}_2^2$, the automorphisms are outer and form the group $PSL(2,2)\cong \mathbb{Z}_6$.   Let us now consider two points of $H_1$, not on the same line; the generated group is $\mbox{Dih}_4$, which is its own automorphism group. We know that $\mbox{Dih}_4$ is the Weyl group of Coxeter system $I_2(4)$. The group generated by an antiflag  
is $\mathbb{Z}_2 \times \mbox{Dih}_4$, corresponding to the Coxeter system $A_1 I_2(4)$; outer automorphisms of the antiflag have the same group structure. The group generated by two lines fail to have a Coxeter structure, neither its automorphism group, but outer automorphisms form the group $S_4 \times \mbox{Dih}_4$, which is the Weyl group of Coxeter system $I_2(4) D_3$.

2) A {\it grid} $H_2$ of $\mbox{GQ}_2$  is of size $3 \times 3$. Its lines have been chosen to carry all the entangled states, i.e., $H_2=\left\{(4,8,12),(9,10,5),(11,6,7),(4,9,11),(8,10,6),(12,5,7)\right\}$. The product of three observables in each of the first three (horizontal) lines is minus the identity matrix, while the product of observables in each of the last three (vertical) lines is the identity matrix. Thus, the grid forms a {\it Mermin square}, which may be used to demonstrate the Kochen-Specker theorem in dimension $4$ \cite{Pauligraphs}. The group generated by a vertical line is the (already encountered) group $\mathbb{Z}_2^2$. The group generated by a horizontal line is $\mathbb{Z}_2^3$, the  automorphisms are outer and form the group $PSL(3,2)\cong PSL(2,7)$ (the group of symmetries of the Klein quartic). The group generated by an antiflag is the (already encountered group) $\mathbb{Z}_2 \times \mbox{Dih}_4$ (the antiflag may contain a line of the horizontal or of the vertical type). Finally, the whole grid generates the group $(\mathbb{Z}_2 \times \mbox{Dih}_4)\rtimes \mathbb{Z}_2$, and the outer automorphisms form the group $(S_3\times S_3)\rtimes \mathbb{Z}_2$. The latter group is not of Coxeter type, but its maximal normal subgroup $S_3\times S_3$ is the Weyl group of Coxeter system $A_2 A_2$. 

3) An {\it ovoid} $H_3$ of $\mbox{GQ}_2$ is a set of five points with no line connecting them (in graph theory, it is called an independent set). Let us take for example $H_3=\left\{1,2,6,9,12\right\}$. The five points belong to a maximal set of five mutually unbiased bases, so that the automorphisms of $H_3$ also define symmetries of mutually unbiased bases. 
Let us denote $m_i$ ($i=1,\ldots,5$) the elements of such a maximal set, one may form groups of increasing size $g_2=\left\langle m_1,m_2\right\rangle$,\ldots, $g_4=\left\langle m_1, m_2,m_3, m_4\right\rangle$ ($g_1$ is the trivial group and $g_5=g_4$). The groups $g_i$ have automorphism groups   $\mbox{Aut}(g_2)=\mbox{Dih}_4$, $\mbox{Aut}(g_3)=\mathbb{Z}_2\times S_4$ and $\mbox{Aut}(g_4)=\mbox{Aut}(g_5)=\mathbb{Z}_2 \wr A_5$, which are Weyl groups of irreducible Coxeter systems of type $I_2(4)$, $B_3$ and $D_5$, respectively (see Table 1). The corresponding outer automorphism groups are $\mathbb{Z}_2$, $\mbox{Dih}_6$ and $S_5$, which are attached to the Coxeter systems $A_1$, $I_2(6)$ and $A_4$.

\subsection{Automorphism groups of mutually unbiased bases for multiple qubit systems}
\label{MUBs}

The finite geometry underlying higher-order qubits was studied in \cite{Pauligraphs, Saniga07, PlanatJPhysA07}. The concept of a $\mbox{GQ}$ generalizes to that of a polar space \cite{Saniga07} but the $n$-qubit spaces ($n>2$) fail to satisfy the axioms of near-linearity \cite{Saniga07}. The latter property may be approached using advanced geometrical concepts such as modules over rings \cite{Saniga08}. Here, we restrict our interest to the geometry underlying mutually unbiased bases, because a link to reflection groups of the unitary type may be observed. Further ramifications between the geometry of symplectic polar spaces and group theory can be found in \cite{Taylor92}.

Let us consider a maximal independent set of the three qubit system as in Sec (\ref{hyper}).
The groups $g_i$, and their automorphism groups $\mbox{Aut}(g_i)$ and $\mbox{Out}(g_i)$, built by increasing the number of generators are given in Table 1.
\begin{table}[h]
\begin{center}
\footnotesize
\begin{tabular}{|r|r|r|r||r|r|}
\hline
$g_i$& $g_2$ & $g_3$ & $g_4$ & $g_5$ & $g_6$ \\
\hline
$G$& $\mathbb{Z}_2^{2}$ & $(\mathbb{Z}_4 \times \mathbb{Z}_2) \rtimes \mathbb{Z}_2$ & $ (\mathbb{Z}_2 \times \mathcal{Q})\rtimes \mathbb{Z}_2$ & $\mathbb{Z}_2 \times((\mathbb{Z}_2 \times \mathcal{Q})\rtimes \mathbb{Z}_2) $ & $g_6$\\
\hline
$\mbox{Aut}(G)$& $\mbox{Dih}_4$ & $\mathbb{Z}_2 \times S_4$ & $\mathbb{Z}_2 \wr A_5$ & $\mathbb{Z}_2^{ 2} \wr A_5$ & $\mathbb{Z}_2^{ 3} \wr A_5$\\
\hline
$\left| \mbox{Aut}(G)\right|$& $8$ & $48$ & $1920$ & $61440$ & $1966080$ \\
\hline
$\mbox{Out}(G)$& $\mathbb{Z}_2$ & $\mbox{Dih}_6$ & $S_5$ & $(\mathbb{Z}_2\times \mathbb{Z}_2) \rtimes M_{20}$ & $(\mathbb{Z}_2\times \mathbb{Z}_4) \rtimes M_{20}^{(2)}$\\
\hline
\end{tabular}
\label{autom}
\normalsize
\caption{Group structure of an independent set of the two-qubit ($g_2$ to $g_4$) and three-qubit systems ($g_2$ to $g_6$). $G$ denotes the identified group and $\mbox{Aut}(G)$ the corresponding automorphism group. $\mathcal{Q}$ is the eight-element quaternion group.}
\end{center}
\end{table} 

Every automorphism group in Table 1 is recognized to be a unitary reflection group of the form $\mathbb{Z}_2^{ l} \wr A_5=G(2^l,2,5)$, the corresponding outer automorphism group is the unitary reflection group $G(2^{l-1},1,5)$ ($l\ge 1$). The latter possesses a factor $M_{20}^{l-1}$ equal to the derived subgroup $G'(2^l,2,5)$, of order $960$ and $15360$, respectively. Group $M_{20}$ (see Appendix 1) is the smallest perfect subgroup for which the derived subgroup is different from the set of commutators; this property applies to group $M_{20}^{(2)}$ and one can surmise that it also applies to higher-order group of the same series. 

The above approach encompasses the automorphisms of some non-additive quantum codes \cite{Rains97}. It also connects to the topological approach of quantum computing as shown in the next section.

\section{Reflection groups, Clifford groups and quantum fault tolerance}
\label{tolerance}
  
In this section, we shall demonstrate that some unitary reflection groups and {\it entangling} Clifford gates \cite{Clark07} are closely related topics.

An $n$-qubit quantum gate can be viewed as a homomorphism from $\mathcal{P}_n$ to itself; in this respect, bijective homomorphisms (automorphisms) are expected to play an important role for protected quantum computations. Clifford gates are a class of group operations stabilizing Pauli operations \cite{Gottesman97, Clark07}. Any action of a Pauli operator $g \in \mathcal{P}_n$ on an $n$-qubit state $\left|\psi\right\rangle$ can be stabilized by a unitary gate $U$ such that $(UgU^{\dag})U\left|\psi\right\rangle=U \left|\psi\right\rangle$, with the condition $UgU^{\dag} \in \mathcal{P}_n$. The $n$-qubit Clifford group (with matrix multiplication for group law) is  
\begin{equation}
\mathcal{C}_n=\left\{U\in U(2^n)|U \mathcal{P}_n U^{\dag}=\mathcal{P}_n\right\}.
\end{equation}
In view of the relation $U^{\dag}=U^{-1}$ for $U \in U(2^n)$, any normal subgroup $\mathcal{Q}_n=\left\{UgU^{-1}, g \in \mathcal{Q}_n, \forall U \in \mathcal{C}_n\right\}$ of $\mathcal{C}_n$ should be useful for stabilizing the errors. A group extension $1\rightarrow\mathcal{Q}_n \rightarrow \mathcal{C}_n \rightarrow \mathcal{C}_n/\mathcal{Q}_n \rightarrow 1$ carries some information about the structure of the error group $\mathcal{P}_n$ and its normalizer  $\mathcal{C}_n $ in $U(2^n)$. Using this strategy, we shall arrive very close to $\mbox{Aut}(\mathcal{P}_n)$, and we shall endow it with a new representation in terms of Clifford gates. 

Our clear-sighted reader will already have noticed that the  dihedral groups $\mbox{Dih}_4$ and $\mbox{Dih}_6$, and the wreath products $\mathbb{Z}_2^l \wr A_5$ encountered in the previous section, are ${\it entangling}$ in the sense of \cite{Clark07} (they contain an entangling gate). Notably, reflection groups $G(2^l,2,5)$ get connected to topological quantum computation {\it \`{a} la Yang-Baxter}, a topic recently investigated in \cite{Kauf04}.

Before handling these {\it topological} gates, we recall the following basic result \cite{Gottesman97}.

Let
H be the Hadamard gate, $P$ the $\pi/4$ phase gate, and let $\mbox{CZ}=\mbox{Diag}(1,1,1,-1)$ be the entangling two-qubit controlled-$Z$ gate. Then any $n$-qubit ($n\ge 2$ ) gate $U$ in $\mathcal{C}_n$ is a circuit involving $H$, $P$ and $\mbox{CZ}$, and conversely.

\subsection{The single qubit Clifford group, $GL(2,3)$ and $G_2$}

The one-qubit Clifford group (No $9$ in the Shephard-Todd list \cite{Shep54,Nebe01}\footnote{The presentation of the Shephard-Todd group No $9$ is $\mathcal{C}_1=\left\langle x_1^2=x_2^2=(x_2^{-1}x_1)^3(x_2x_1)^3=1\right\rangle$.})  possesses a representation in terms of the gates $H$ and $P$ as $\mathcal{C}_1=\left\langle H,P\right\rangle$. Its order is $\left|\mathcal{C}_1\right|=192$. The center is $Z(\mathcal{C}_1) \cong \mathbb{Z}_8$, the central quotient is $\tilde{\mathcal{C}_1}=S_4$ and the commutator subgroup is $\mathcal{C}_1'\cong SL(2,3)$.

Let us display two important split extensions. One is related to the {\it magic} group $\left\langle T,H\right\rangle\cong GL(2,3)$, where $T=\exp(i\pi/4)PH$, which was introduced in \cite{Kitaev05} 
\begin{equation}
1 \rightarrow GL(2,3) \rightarrow \mathcal{C}_1 \rightarrow \mathbb{Z}_4 \rightarrow 1. 
\end{equation}
A second important split extension sends back to the reflection group $\mbox{Dih}_6$ encountered in Eq (\ref{Out2})
\begin{equation}
1 \rightarrow \mathcal{P}_1 \rightarrow \mathcal{C}_1 \rightarrow \mbox{Dih}_6 \rightarrow 1. 
\label{single}
\end{equation}
The Clifford group $\mathcal{C}_1$ {\it modulo} the Pauli group $\mathcal{P}_1$ corresponds to the outer automorphism group of $\mathcal{P}_1$ (the word {\it modulo} means that we are dealing with the group quotient $\mathcal{C}_1/\mathcal{P}_1$). This interesting outcome (relating issues about the outer automorphism group of the Pauli group and issues about the quantum gates, via the entangling dihedral group $\mbox{Dih}_6$ of the $G_2$ Coxeter system)  turns out to still hold for the two-qubit system. 

\subsection{The two-qubit Clifford group, $U_6$ and $A_1A_5$}

As for the two-qubit Clifford group\footnote{The Clifford group $\mathcal{C}_2$ contains three maximal normal subgroups of order $46080$. One of them is the reflection group No $31$ in the Shephard-Todd list \cite{Nebe01}. Its presentation is $\langle x_1^2=x_2^2=x_3^2=x_4^2=x_5^2=(x_1x_4)^2=(x_2x_4)^2=(x_2x_5)^2=(x_2x_1)^3=(x_3x_2)^3=(x_4x_3)^3=(x_5x_4)^3=$ $x_5x_1x_3x_1x_5x_3=x_1x_5x_3x_1x_3x_5=1 \rangle$.}, the representation is $\left\langle \mathcal{C}_1 \otimes \mathcal{C}_1, \mbox{CZ}\right\rangle$.
 One has $\left|\mathcal{C}_2\right|=92160$ and $Z(\mathcal{C}_2)=Z(\mathcal{C}_1)$. The central quotient $\tilde{\mathcal{C}}_2$ satisfies
\begin{equation}
1 \rightarrow U_6 \rightarrow \tilde {\mathcal{C}}_2 \rightarrow \mathbb{Z}_2 \rightarrow 1. 
\label{V6}
\end{equation}
The group $U_6=\mathbb{Z}_2^4 \rtimes A_6$ was found in Eq (\ref{U6}) to be the stabilizer of a hexad in $M_{22}$. Group $\tilde{\mathcal{C}}_2$ is twice
 larger than $\mbox{Aut}(\mathcal{P}_2)$ but both possess $U_6$ as an extension group. Another relevant expression is the Clifford group $\mathcal{C}_2$ {\it modulo} the Pauli group $\mathcal{P}_2$ as the direct product  
\begin{equation}
\mathcal{C}_2/\mathcal{P}_2=\mathbb{Z}_2 \times S_6,
\label{two}
\end{equation}
a group also isomorphic to $\mbox{Out}(\mathcal{P}_2)$, as found in Eq \ref{Out1}. The reducible Coxeter system $A_1A_5$ underlies these group isomorphisms. 
Another relevant isomorphism is $S_6 \cong \mbox{Sp}(4,2)$. The symplectic groups $\mbox{Sp}(2n,2)$ are well known to control the symmetries of $n$-qubit Clifford groups \cite{Calder98,Vourdas07}.
 
\subsection{The three-qubit Clifford group and $E_7$}

To generate the three-qubit Clifford group, one can use the representation $\mathcal{C}_3=\left\langle H\otimes H \otimes P,H \otimes \mbox{CZ},\mbox{CZ} \otimes H\right\rangle$. The following semi-direct product is known \cite{Calder98} 
\begin{equation}
\tilde{\mathcal{C}}_3 \cong\mathbb{Z}_2^6 \rtimes W'(E_7),~~\mbox{with}~W'(E_7)\cong \mbox{Sp}(6,2).
\label{E7}
\end{equation}
One has $\left|Z(\mathcal{C}_3)\right|=8$ and $\left|\tilde{\mathcal{C}}_3\right|=92~897~280$. It should be clear that, when one passes from two to three qubits, the Weyl group $W'(E_7)\cong \mbox{Sp}(6,2)$ replaces $W(A_5)=S_6$.
Based on cardinalities, one can suspect that a relation, generalizing (\ref{single}) and (\ref{two}), relating the outer automorphism group and $\mathcal{C}_3$ {\it modulo} $\mathcal{P}_3$ still holds, i.e., $\mathcal{C}_3/\mathcal{P}_3=\mbox{Out}(\mathcal{P}_3)=\mathbb{Z}_2 \times \mbox{Sp}(6,2)$. This relation suggests the possible existence of irreducible Coxeter systems {\it hidden} in $\mathcal{C}_2$ and $\mathcal{C}_3$, that would play a similar role as the Weyl group $G_2$ of the hexagon plays for the single qubit system. This hypothetical system can be foreseen by reading Sec \ref{MUBs} and will be uncovered in the next section.   

\subsection{Topological entanglement, the Yang-Baxter equation, the Bell groups and Coxeter system $E_6$}
\label{topological}

Topological quantum computing based on anyons was proposed as a way of encoding quantum bits in nonlocal observables that are immune of decoherence \cite{Kitaev04,Preskill98}. The basic idea is to use pairs $\left|v,v^{-1}\right\rangle$ of \lq\lq magnetic fluxes" for representing the qubits and permuting them within some large enough nonabelian finite group $G$ such as $A_5$. The \lq\lq magnetic flux" carried by the (anyonic) quantum particle is labeled by an element of $G$, and \lq\lq electric charges" are labeled by irreducible representation of $G$ \cite{Ogburn}.

The exchange within $G$ modifies the quantum numbers of the fluxons according to the fundamental logical operation
\begin{equation}
 \left|v_1,v_2\right\rangle \rightarrow \left|v_2,v_2^{-1} v_1 v_2\right\rangle,
\nonumber
\end{equation}
a form of Aharonov-Bohm interactions, which is nontrivial in a nonabelian group. This process can be shown to produce universal quantum computation. It is intimately related to topological entanglement, the braid group and unitary solutions of the Yang-Baxter equation \cite{Kauf04}
\begin{equation}
(R \otimes I)(I \otimes R)(R \otimes I)=(I \otimes R)(R \otimes I)(I \otimes R),\nonumber
\end{equation}
in which $I$ denotes the identity transformation and the operator $R$: $V \otimes V \rightarrow V \otimes V$ acts on the tensor product of the two-dimensional vector space $V$. One elegant unitary solution of the Yang-Baxter equation is a universal quantum gate known as the Bell basis change matrix
\begin{equation}
R=1/\sqrt{2}\left(\begin{array}{cccc} 1 & 0 & 0 & 1 \\0 & 1	& -1 & 0 \\ 0 & 1 & 1 & 0 \\-1 & 0 & 0 & 1\\ \end{array}\right)\nonumber.
\end{equation}
This gate is an {\it entangling} \cite{Clark07} and also a {\it match} \cite{Jozsa08} gate. In the words of \cite{Kauf04}, matrix $R$ \lq\lq can be regarded as representing an elementary bit of braiding represented by one string crossing over one another".  In this section, we shall not examine further the relation to the braid group, but explore the relation of the gate $R$ to unitary reflection groups such as $D_5$ and higher-order systems such as those encountered in Sec \ref{MUBs}. See also \cite{PlanatBN} for more advanced topics related to the Bell groups.

This can be done by replacing the gate $\mbox{CZ}$ in the definition of the Clifford group by the new entangling gate $R$ and by building the {\it Bell group} as follows
\begin{equation}
\mathcal{B}_2=\left\langle H \otimes H, H \otimes P, R\right\rangle.
\end{equation}
The Bell group  $\mathcal{B}_2$ is a non-normal subgroup of $\mathcal{C}_2$. It presents a structure quite similar to $\mathcal{C}_2$: the central quotient, as the one of its parent, only contains two normal subgroups $\mathbb{Z}_2^4$ and $M_{20}=\mathbb{Z}_2^4 \rtimes A_5$ (The alternating group $A_5$ replaces $A_6$, and $M_{20}$ replaces $U_6$ of Eq \ref{V6}.) The new important feature is that $\tilde{\mathcal{B}}_2$ involves the Weyl group of the irreducible Coxeter system $D_5$, already encountered in the automorphisms of a complete set of mutually unbiased bases. The central quotient $\tilde{\mathcal{B}}_2$ reads

\begin{equation}
\tilde{\mathcal{B}}_2 \cong \mathbb{Z}_2^4 \rtimes S_5=W(D_5).
\label{D5}
\end{equation}

The Pauli group $\mathcal{P}_2$ is normal in $\mathcal{B}_2$ and a relation similar to (\ref{two}) holds

\begin{equation}
\mathcal{B}_2/\mathcal{P}_2 \cong \mathbb{Z}_2 \times S_5.
\label{newtwo}
\end{equation}

Let us pass to the generalization of $\mathcal{B}_2$ to the three qubit Bell group
\begin{equation}
\mathcal{B}_3=\left\langle H\otimes H \otimes P,H \otimes \mbox{R},\mbox{R} \otimes H\right\rangle.
\end{equation}
Now, $\mathcal{B}_3$ is a non-normal subgroup of the three qubit Clifford group $\mathcal{C}_3$. Its central quotient may be written in a form replacing (\ref{E7})
\begin{equation}
\tilde{\mathcal{B}}_3\cong \mathbb{Z}_2^6 \rtimes W'(E_6),~~\mbox{with}~W'(E_6) \cong \mbox{SU}(4,2)\cong \mbox{PSp}(4,3),
\label{E6}
\end{equation}
in which $SU(4,2):=SU(4,\mathbb{F}_2)$, the special unitary group of four by four (determinant one) matrices over the field $\mathbb{F}_2$, is isomorphic to the projective symplectic group $SU(4,3):=\mbox{PSp}(4,\mathbb{F}_3)$ over the field $\mathbb{F}_3$.

The (exceptional) irreducible Coxeter system $E_6$ controls the structure of the Bell group $\mathcal{B}_3$. The Coxeter system is of rank six and the generating relations
  are
\begin{eqnarray}
& x_1^2=x_2^2=\ldots=x_6^2= \nonumber\\
&(x_1x_2)^2=(x_2x_3)^2=(x_1x_4)^2=(x_1x_5)^2=(x_2x_5)^2= \nonumber \\
&(x_3x_5)^2=(x_1x_6)^2=\ldots (x_4x_6)^2 \nonumber \\
&(x_3x_1)^3=(x_4x_2)^3=(x_4x_3)^3=(x_5x_4)^3=(x_6x_5)^3=1.
\label{presE6}
\end{eqnarray}
The weight lattice of the Weyl group $W(E_6)$ is as follows  

$$\mathcal{L}_{W(E_6)}:=\left(\begin{array}{cccccc} 4 & 3 & 5 & 6 & 4 & 2 \\3	 & 6 & 6 & 9 & 6 & 3 \\ 5 &  6 & 10 & 12 & 8 & 4 \\
                              6 & 9 & 12&18 & 12& 6 \\4  & 6 & 8 & 12&10 & 5 \\ 2 & 3 &  4  & 6 & 5 & 4   \end{array}\right)$$

%
%

The Weyl group $W(E_6)$, of order $51840$, stabilizes the $E_6$ polytope discovered in 1900 by T. Gosset.  
The isomorphism of $W'(E_6)$ to $SU(4,2)$ indicates a link of the three-qubit  Pauli group to the generalized quadrangle $\mbox{GQ}_3$ of the symplectic geometry of dimension $4$ over the field $\mathbb{F}_3$ (see \cite{Taylor92}, p 125). This generalizes our result concerning the symplectic generalized quadrangle $\mbox{GQ}_2$ associated with the two-qubit Pauli group $\mathcal{P}_2$. The isomorphism of $W'(E_6)$ to the groups $SU(4,2)$ and  $\mbox{PSp}(4,3)$ provides an example of a group with two different $\mbox{BN}$ pair structures (see \cite{Taylor92} and \cite{PlanatBN} for the meaning of this group structure).

\section{Discussion}

Reflection groups form the backbone of the representation theory of Lie groups and Lie algebras, which were proposed by E.~P. Wigner  through a study of the Poincar\'{e} group to understand the space-time symmetries of elementary particles. In this essay, we have unraveled specific symmetries of multiple qubit systems (sets of $\frac{1}{2}$-spin particles) and found them to be governed by specific Coxeter systems (such as $D_5$ and $E_6$) and complex reflection groups (such as $G(2^l,2,5)$). These symmetries have particular relevance to the topological approach of quantum computation \cite{Kauf04} and to entangling groups of quantum gates \cite{Clark07}.

%
%

We would like to view  the reflection groups $W(D_5)$ and $W(E_6)$ as well as the associated central quotient of Bell groups $\tilde{\mathcal{B}}_2=\mathbb{Z}_2^4 \rtimes S_5=W(D_5)$ (see Eq (\ref{D5})) and $\tilde{\mathcal{B}}_3=\mathbb{Z}_2^6 \rtimes W'(E_6)$ (see Eq (\ref{E6})) in an unifying geometrical perspective. Let us start by looking at the list of maximal subgroups of $W(E_6)$. One recovers $W'(E_6)$ (order $25920$ and length $1$), $W(D_5)$ (order $1920$ and index $27$), $W(F_4)$ (order $1152$ and index $45$) and $A_6.\mathcal{Z}_2^2$ (order $1440$, index $36$). These numbers are akin to the structure of smooth cubic surfaces.

A {\it smooth cubic surface} $\mathcal{K}_3$ of the complex three-dimensional projective space contains a maximum of $27$ lines in general position 
. This results goes back to the middle of 19th century with contributions by A. Cayley, L. Cremona and many others \cite{Dolga04,Hunt}. One can find sets of six mutually {\it skew} lines, and very special arrangements of {\it Schl\"{a}fli's double sixes} of lines (whose incidence is nothing but a $6 \times 6$ grid with the points of the diagonal missing). One can also form configurations of {\it tritangent planes}, i.e., planes that intersect the surface along the union of three lines. The symmetry group of the configuration of the $27$ lines on $\mathcal{K}_3$ is $W(E_6)$, the stabilizer of a line on the cubic surface is $W(D_5)$ \cite{Colo05} and the ratio of cardinalities is $|W(E_6)|/|W(D_5)|=27$. Each of the $45$ tritangent planes is stabilized by the Weyl group $W(F_4)$, and each of the $36$ double sixes possesses the non-split product $A_6.Z_2^2$ as group of automorphisms (see also Eq (\ref{U6})). Thus, the geometrical structure of $\mathcal{K}_3$ perfectly fits the structure of $W(E_6)$ into its non-solvable maximal subgroups. See also \cite{Manivel,Allcock02}.

Another stimulating topic concerns the entangled component of the three-qubit Clifford group. The group $\mathcal{C}_3$ contains simple subgroups of order $168$, $12096$ and $6048$, that one may identify to the simple groups of Lie type $A_2(2)=PSL(2,7)$, $G_2(2)$ and $G_2(2)'$, respectively. The group $A_2(2)$ was already encountered in Sec 3.4 as the automorphism group of an entangling triple of operators. It is the smallest Hurwitz group, with presentation $\left\langle x,y|x^2=y^2=(xy)^7=[x,y]^4=1\right\rangle$ \cite{Conder90}. The smallest exceptional Lie group $G_2(2)$ can be seen as the automorphism group of the octonions or as the automorphism group of the split Cayley hexagon of order two which was recently found to underlie the observables of the three-qubit system \cite{Levay08} \footnote{The representation found for the group $A_2(2)$ is different from \cite{Levay08}. The generators $x_1$ and $x_2$ contain the Pauli spin matrices and satisfy  $x_1^2=x_2^4=(x_1x_2^{-1})^7=(x_2^{-2}x_1)^2x_2^2x_1=1$, where

$x_1=\frac{1}{2}\left(\begin{array}{cccc} \sigma_0 & -\sigma_z & i\sigma_z &-i\sigma_0 \\-\sigma_z & \sigma_0 & i\sigma_0 & -i \sigma_z \\ -i \sigma_z & -i\sigma_0 & \sigma_0 & \sigma_z \\ i \sigma_0 & i\sigma_z & \sigma_z & \sigma_0 \end{array}\right)$, $x_2^2=\frac{1}{2}\left(\begin{array}{cccc} \sigma_0 & \sigma_y & i\sigma_y &-i\sigma_0 \\ \sigma_y & \sigma_0 & -i\sigma_0 & i \sigma_y \\ -i \sigma_y & i\sigma_0 & \sigma_0 & \sigma_y \\ i \sigma_0 &- i\sigma_y & \sigma_y & \sigma_0 \end{array}\right)$.}.

 As a final note, the quest for fault tolerance in quantum computing seems to lead to intriguing relationships between several areas (group theory, algebraic geometry and string theory) so far not fully explored.

\section*{Acknowledgements}

The first author acknowledges the feedback obtained by Robert Raussendorf, Miguel Angel Martin-Delgado, Richard Jozsa, Metod Saniga and Patrick Sol\'{e}. This research was partially supported by Perimeter Institute for Theoretical Physics. Research at Perimeter Institute is supported by the Government of Canada through Industry Canada and by the Province of Ontario through the Ministry of Research \& Innovation.

\section*{Appendix 1}

\subsection*{On group commutators and group extensions}
An on-line introduction to group theory may be found in Ref \cite{Milne}.

A {\it normal subgroup} $N$ of a group $G$ is invariant under conjugation: that is, for each $n$ in $N$ and each $g$ in $G$, the conjugate element $g n g^{-1}$ still belongs to $N$. Noticeable examples are as follows. The center $Z(G)$ of a group $G$ (the set of all elements in $G$, which commute with each element of $G$) is a normal subgroup of $G$. The group $\tilde{G}=G/Z(G)$ is called the central quotient of $G$. Our second example is the subgroup $G'$ of commutators (also called the derived subgroup of $G$). It is the subgroup generated by all the commutators $[g,h]=ghg^{-1}h^{-1}$ of elements of $G$. The set $K(G)$ of all commutators of a group $G$ may depart from $G'$ \cite{kappe}. 

Normal subgroups are the cornerstone of {\it group extensions}. Let $\mathcal{P}$ and $\mathcal{C}$ be two groups such that $\mathcal{P}$ is normal subgroup of $\mathcal{C}$. The group $\mathcal{C}$ is an extension of $\mathcal{P}$ by $H$ if there exists a short exact sequence of groups
\begin{equation}
1 \rightarrow \mathcal{P} \stackrel{f_1}{\rightarrow} \mathcal{C} \stackrel{f_2}\rightarrow H \rightarrow 1,
\nonumber
\end{equation}
in which $1$ is the trivial (single element) group.

The above definition can be reformulated as: (i) $\mathcal{P}$ is isomorphic to a normal subgroup $N$ of $\mathcal{C}$, (ii) $H$ is isomorphic to the quotient group $\mathcal{C}/N$.

In an exact sequence the image of $f_1$ equals the kernel of $f_2$; it follows that the map $f_1$ is injective and $f_2$ is surjective. 

Given any groups $\mathcal{P}$ and $H$ the {\it direct product} of $\mathcal{P}$ and $H$ is an extension of $\mathcal{P}$ by $H$.

The {\it semidirect product} $\mathcal{P} \rtimes H$ of $\mathcal{P}$ and $H$ is as follows.  The group $\mathcal{C}$ is an extension of $\mathcal{P}$ by $H$ (one identifies $\mathcal{P}$ with  a normal subgroup of $\mathcal{C}$) and: (i) $H$ is isomorphic to a subgroup of $\mathcal{C}$, (ii) $\mathcal{C}$=$\mathcal{P} H$ and (iii) $\mathcal{P}\cap H=\left\langle 1\right\rangle$. One says that the short exact sequence splits.   

The {\it wreath product} $M \wr H$ of a group $M$ with a permutation group $H$ acting on $n$ points is the semidirect product of the normal subgroup $M^n$ with the group $H$, which acts on $M^n$ by permuting its components.

Let $G=\mathbb{Z}_2 \wr A_5$, in which $A_5$ is the alternating group on five letters, then $G'$ is a perfect group with order $960$ and  one has $G' \ne K(G)$. Let $H=Z_2^5 \rtimes A_5$, one can think of $A_5$ having a wreath action on $Z_2^5$. The group  $G'=\tilde{H}=M_{20}$  \cite{Brauer} is the smallest perfect group having its commutator subgroup distinct from the set of the commutators \cite{kappe}. Some unitary reflection groups (see Sec \ref{MUBs}) specify wreath actions in an essential way, seeing that $G(2^l,2,5)= \mathbb{Z}_2^l \wr A_5$.

\subsection*{On group of automorphisms}

Given the group operation $\ast$ of a group $G$, a group endomorphism is a function $\phi$ from $G$ to itself such that $\phi(g_1 \ast g_2)= \phi(g_1) \ast \phi(g_2)$, for all $g_1,g_2$ in $G$. If it is bijective, it is called an {\it automorphism}. An automorphism of $G$ that is induced by conjugation of some $g\in G$ is called {\it inner}. Otherwise it is called an {\it outer} automorphism. Under composition the set of all automorphisms defines a group denoted $\mbox{Aut}(G)$. The inner automorphisms form a normal subgroup $\mbox{Inn}(G)$ of $\mbox{Aut}(G)$, that is isomorphic to the central quotient of $G$. The quotient $\mbox{Out}(G)=\mbox{Aut}(G)/\mbox{Inn}(G)$ is called the outer automorphism group.

\subsection*{On maximal non-solvable subgroups}

A subgroup $H$ of $G$ is said to be a {\it maximal} subgroup of $G$ if $H\neq G$ and there is no subgroup $K$ of $G$ such that $H<K<G$. A normal subroup $N$ of $G$ is a maximal normal subgroup iff the quotient $G/N$ is a {\it simple group} (By definition a simple group $G$ only contains the normal subgroups $\left\{1\right\}$ and $G$ itself).

Let $H$ a subgroup of $G$, and let $G=G_0\triangleleft G_1 \triangleleft \cdots \triangleleft G_n=H$ be a series of subgroups with each $G_i$ a normal subgroup of the previous one $G_{i-1}$. A group $G$ is said to be {\it solvable} if the series ever reaches the trivial subgroup $\left\{1\right\}$ and all the quotient groups $G_i/G_{i+1}$ are abelian. An equivalent definition is that every subgroup of the series is the commutator subgroup of the previous one.
Otherwise  $G$ is called a {\it non-solvable} group.

Non-solvable maximal subgroups of the reflection group $W(E_6)$ have a geometrical significance displayed in the conclusion of the present paper.


\end{document}